\documentclass[twocolumn]{aastex6}

  \usepackage{amsmath}
  \usepackage{tabularx}
  \usepackage{color}
  \usepackage{graphicx}
  \usepackage{rotating}
  \usepackage{afterpage}
\usepackage{longtable}
\begin{document}

\title{A cloud-based architecture for the Cherenkov Telescope Array observation simulations. \\
Optimisation, design, and results} 

\author{M.\ Landoni\altaffilmark{1},  
P.\ Romano\altaffilmark{1}, 
S.\ Vercellone\altaffilmark{1}, 
J. \ Kn{\"o}dlseder\altaffilmark{2}, 
A.\ Bianco\altaffilmark{1},  
F.\ Tavecchio\altaffilmark{1}, 
A.\ Corina\altaffilmark{3}
}
\altaffiltext{1}{INAF, Osservatorio Astronomico di Brera, Via E. Bianchi 46 I-23807 Merate (LC),  ITALY}
\altaffiltext{2}{Institut de Recherche en Astrophysique et Planetologie, 9 avenue Colonel-Roche, 
                   31028, Toulouse Cedex 4, France}
\altaffiltext{3}{Amazon Web Services Inc. - P.O. Box 81226, Seattle, WA 98108-1226}

\begin{abstract}
Simulating and analysing detailed observations of astrophysical sources for very high energy (VHE) experiments, 
like the Cherenkov Telescope Array (CTA), can be a demanding task especially in terms of CPU 
consumption and required storage. 
In this context, we propose an innovative cloud computing architecture based on Amazon Web Services (AWS)
aiming to decrease the amount of time required to simulate and analyse a given field 
by distributing the workload and exploiting the 
large computational power offered by AWS. We detail how the various services offered by the Amazon 
online platform are jointly used in our architecture and we report a comparison of the execution times 
required for simulating observations of a test source with the CTA, 
by a single machine and the cloud-based approach. 
We find that, by using AWS, we can run our simulations more than 2 orders of magnitude faster 
than by using a general purpose workstation for the same cost.  
We suggest to consider this method when observations need to be simulated, analysed, and concluded 
within short timescales.

\end{abstract}

\section{Introduction}

 While the very high energy (VHE, above a few tens of GeV) $\gamma$-rays astrophysics community is 
obtaining astounding results with  the  analysis of  data from the current generation of 
ground-based imaging atmospheric Cherenkov telescopes 
\citep[IACTs, see][]{2009ARA&A..47..523H,lemoine}, 
a new generation of IACTs is already being developed. 
The Cherenkov telescope array (CTA) has been proposed \citep[][]{2011ExA....32..193A,2013APh....43....3A}
to dramatically boost the current IACT performance and to increase the breadth and depth of the VHE science. 

%
 To obtain the required wide energy range covered by CTA (from 20 GeV up to 300\,TeV) 
the array will be composed of different classes of telescopes, namely, 
the large-sized telescopes (LSTs, D$\sim23$\,m), 
   which will lower the energy threshold down to a few tens of GeV, 
medium-sized telescopes  (MSTs, D$\sim12$\,m) 
  which will improve the sensitivity in the 0.15--5\,TeV energy range by a factor of five-to-ten, 
and small-sized telescopes (SSTs, primary mirror D$\sim4$\,m)  
  from which the study of the Galactic plane in the energy range beyond 100 TeV will benefit the most. 
Furthermore, the full array will be installed in two sites, one for each hemisphere to allow an 
all-sky coverage. 
The baseline setup currently includes ~\citep[][]{2017AIPC.1792b0014H,2017Msngr.168...21H} 
4 LSTs and 15 MSTs in Northern site, covering an area of $\sim1$\,km${^2}$, 
at the Observatorio del Roque de los Muchachos on the island of La Palma (Spain), and 
4 LSTs, 25 MSTs, and 70 SSTs in Southern site, covering an area of about 4\,km$^{2}$, 
at the European Southern Observatory’s (ESO’s) Paranal Observatory in the Atacama Desert (Chile).

CTA is currently in the scientific assessment phase of simulating feasibility and 
scientific return of potential astrophysical targets which, in turn, can be used to determine future
observing plans that maximise the overall payoff  along the whole CTA lifetime.  
This often implies episodic, highly CPU-intensive simulations that are performed on 
specific science projects within broader topics on very short timescales. 
Under these conditions, it is generally not cost effective to purchase, set up, and maintain 
a large enough cluster of computers to perform the task, and it may be cheaper to buy CPU time 
(and all correlated services of moving and storing large amounts of data) 
in a cloud platform \citep[see e.g.][]{Williams2018}.

In order to use cloud computing, two steps need to be taken. 
The first is to choose a cloud platform among the many currently available 
(e.g.\ Amazon Web Services, Google Cloud Platform), 
which offers the flexibility of a solution tailored to everyone's computing and storage requirements, 
that can also meet their financial constraints. 
The second step is to adapt all software that needs to be run for the simulations 
so they can run in a parallel fashion (i.e., able to exploit many cores on a machine)  
and to be distributed, thus sharing workload among many computers across the network. 

For the purpose of our simulations for CTA we chose Amazon Web Services
(AWS)\footnote{\href{https://aws.amazon.com/}{https://aws.amazon.com/}},  
thanks to a combination of high reliability, low cost, ease of service, 
and previous positive experience \citep{genonihires}.  
We also used the Docker platform\footnote{\href{https://www.docker.com/}{https://www.docker.com/}} 
to ease the distribution of the tasks to run.  

To perform the case study we present in this work, we used {\tt ctools} 
\citep[][v.\ 1.4.2]{Gammalib_ctools_2016}\footnote{\href{http://cta.irap.omp.eu/ctools/}{http://cta.irap.omp.eu/ctools/}}, 
an analysis package for IACT data. 
{\tt ctools}  is not designed to be operated parallely and distributed across many different nodes. For this reason we ran each simulation as a sequence of {\tt ctools} tasks, executed through {\tt shell} scripts, 
independently in a single thread without message passing between processes throughout the network. Final results are stored at the end of computation in the local file system of each node.

In this paper we show an example of an extensive set of simulations 
based on a Monte-Carlo sampling of the CTA Instrument Response functions of test
astrophysical sources as seen through the eyes of the forthcoming CTA array. We briefly describe the tasks executed to perform our simulations, and 
the requirements for their parallelisation and distribution  
(Sect.~\ref{methods:sims}). 
We then describe our novel approach to running these simulations 
based on cloud computing (Sect.~\ref{methods:architecture}). 
Finally, the results and the advantages are discussed in terms of optimisation of computing times 
using  cloud computing for this kind of work in Sect.~\ref{methods:results} and 
~\ref{methods:budget}.

 	 \section{A case study } \label{methods:sims}

%
  \begin{table}[t] 	
 \tabcolsep 2pt  
\begin{center} 	
 \caption{Array of {\tt ctools} simulations. All realizations were obtained considering only the Southern CTA site and a zenith angle of 20\,deg. 	
 \label{methods:tab:crab} }	
 \begin{tabular}{lllllll}	
 \hline 
 \hline 
 \noalign{\smallskip} 
  Model                        & IRF	                                                & Expo.	  & Sim.   &   CPU     \\
                                   & 	                                                        & 	(h)      & $N$   &   Time\tablenotemark{a}    \\
 \noalign{\smallskip} 
 \hline 
 \noalign{\smallskip} 
Crab			   &	{\tt South\_z20\_average\_5h}	&	5	 &	1000&  8$^{\rm h}$   48$^{\rm m}$  \\
0.1Crab			   &	{\tt South\_z20\_average\_5h}	&	5	 &	1000&  8$^{\rm h}$     6$^{\rm m}$ \\
0.01Crab                    &	{\tt South\_z20\_average\_5}h	&	5	 &	1000&  9$^{\rm h}$ 33$^{\rm m}$ \\
mCrab\tablenotemark{b}  &	{\tt South\_z20\_average\_5h}	&	5	 &	1000&  24$^{\rm h}$ 58$^{\rm m}$ \\    
mCrab\tablenotemark{b,c}&	{\tt South\_z20\_average\_5h}	&	10	 &	1000&  43$^{\rm h}$ 10$^{\rm m}$  \\   
mCrab			   &	{\tt South\_z20\_average\_50h}	&	50	 &	1000&  74$^{\rm h}$ 27$^{\rm m}$  \\    
  \noalign{\smallskip}
  \hline
  \end{tabular}
\tablenotetext{a}{Run time for 1000 realizations (single core). }
\tablenotetext{b}{The source is not detected in most of the realizations
(see  Fig.~\ref{methods:fig:mCrab_over_TS} and Table~\ref{methods:tab:results}). }
\tablenotetext{c}{We used the  IRFs relative to the closest exposure. } 
  \end{center}
  \end{table} 
As a case study we considered 4 test, 
point-like sources with a simple Crab-like power-law spectrum spanning 4 orders of magnitude 
in flux (1, 0.1, 0.01, and 0.001 Crab). These sources can be used to estimate the run times for more realistic sources,
as we discuss in Sect.~\ref{methods:results}. 
%
%
Our simulations were performed with the {\tt ctools}  
package and the public CTA instrument response 
files\footnote{\href{https://www.cta-observatory.org/science/cta-performance/}{https://www.cta-observatory.org/science/cta-performance/}  }
(IRF, v.\ prod3b-v1). 

To define the spectrum of our test sources we adopted a power-law spectral model, 
described  as a monochromatic flux  
\begin{equation}
M_{\rm spectral}(E)=k_0 \left( \frac{E}{E_0} \right) ^{\gamma} ,
\end{equation} 
where $k_0$ is the normalisation (or Prefactor, in units of ph\,cm$^{-2}$\,s$^{-1}$\,MeV$^{-1}$), 
 $E_0$ is the pivot energy in MeV\footnote{Generally fixed, at $10^{6}$\,MeV.}, and   
$\gamma$ is the power-law photon index. 
A Crab-like power-law spectrum is therefore described by 
$k_0=5.7\times10^{-16}$\,ph\,cm$^{-2}$\,s$^{-1}$\,MeV$^{-1}$, and 
$\gamma=-2.48$, so that in the 0.1--100\,TeV band 
1 Crab $=5.248\times10^{-10}$\,erg\,cm$^{-2}$\,s$^{-1}$.   
We placed our test sources at the coordinates of a known galaxy, NGC~1068, 
RA(J$2000)=02^{\rm h}\, 42^{\rm m}\, 40\fs70$, 
Dec(J$2000)=-00^{\circ}\, 00^{\prime}\, 48\farcs0$, 
so that they are visible from both CTA sites at a zenith angle of 20\,deg. 
For purposes of method validation and calculation of run times, 
we only consider the southern site because it provides a wider energy coverage.  
We therefore chose our IRFs based on coordinates and required exposure time, 
as reported in Table~\ref{methods:tab:crab},  Col.~2.

For the residual cosmic-ray background we included the instrumental background described in the IRFs 
and no further contaminating astrophysical sources in the 
5\,deg field of view (FOV) were considered for event extraction.

We define 
a ``simulation'' as a set of $N$ independent realizations. 
Each realization is performed through a {\tt shell} script driving a sequential series of commands,  
alternating purely astrophysical computations performed through  {\tt ctools} tasks,
and housekeeping scripts.  
In our specific case, 
a realization includes first running the task {\tt ctobssim} within {\tt ctools} 
to create one event list based on our input model, including 
background events that were randomly drawn from the background model that is shipped with the IRFs. 
The randomisation is controlled by a seed that is unique to this realization. 
Subsequently,  the task {\tt ctlike} reads in the event file and the input model file and,  
by using an unbinned maximum likelihood model fitting,  
determines the best fit spectral parameters from which we derive the flux, 
as well as the covariance matrices and the test statistics (TS) value \citep[][]{Mattox1996:Egret}. 

To overcome the impact of a given statistical realization on the fit results, 
we performed for each spectral model sets of 
$N=1000$ statistically independent 
realizations by changing the {\tt ctobssim} seed value, 
thus calculating 1000 sets of each spectral parameter and TS. 
The value of each spectral parameter and its uncertainty  
are then calculated as the mean and standard deviation drawn from the distribution 
of the values of such parameter. 

In our case, each simulation is described by a ``simulation table'', i.e., 
a list of $N=1000$ calls to a script that will perform one realization. 
For the simulation table in order to run in a parallel fashion, 
each step needs to be uniquely dependent on the randomization seed, 
and to rely on uniquely defined variables and input/output files.

A summary of the simulations is shown in Table~\ref{methods:tab:crab}, 
where, for each test source, 
Col.~2 reports the IRF chosen, 
Col.~3 the exposure, 
Col.~4  the number of realizations run for each simulation,
and Col.~5 the typical run time for 1000 realizations. We used an  
Intel\textsuperscript{\textregistered}  Xeon\textsuperscript{\textregistered}  CPU E5-2620 v4 @ 2.10GHz machine (8 cores) with 82GB RAM, 
running RedHat 7.5 with gcc compiler v.4.8.5, 
for all simulations. 

 	 \section{Living on a Cloud } \label{methods:architecture}

The key to increase by orders of magnitude the performance of trivially parallel simulations, where no sustained network communication between processes is required, resides 
in the parallelisation and distribution of the workload across a cluster 
of computers spread out through the network.
This could be reached by using an arbitrarily large number of computers 
with parallel use of CPUs on each node. 
In practice, a compromise between overall run-time and cost (the faster, the more expensive) 
must be reached depending on the goals and time-frame of the project and its funding. 
In what follows we describe our solution and we briefly review the main services used to achieve our goal.

\begin{figure}[t] 
\hspace{-0.20truecm}
\includegraphics[scale=0.32,angle=0]{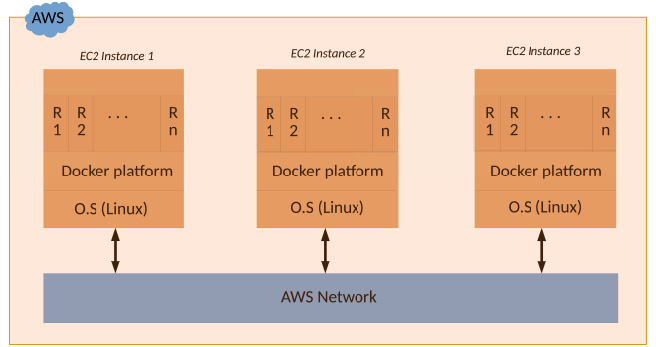}
\caption{AWS EC2 method for the distribution of the workload: the {\tt ctools} suite is 
containerised through a {\tt Docker} in each instance. 
Each instance allows a large number of realizations to be run simultaneously. 
} 
\label{methods:fig:awsec2} 
\end{figure} 

\subsection{The Docker Platform } \label{methods:docker} 

The first step of the distribution of the simulations is to make a fully functional image of the software  
that can be duplicated indefinitely on any computer.   
We reached this goal with
{\tt Docker}\footnote{\href{https://www.docker.com/}{https://www.docker.com/}. }, 
a software program that allowed us to ``containerise'' into one single virtual executable file 
(called Docker Image) all relevant software, including libraries, environment configurations, and software installation,
which is fully portable from one computer to another in a complete platform-independent fashion. 
Furthermore, while {\tt Docker} works like a virtual machine, it is nonetheless orders of magnitude lighter 
than the latter. This allows the simultaneous execution of dozen of runs, called ``containers'',  of the executable described above 
on the same machine while guaranteeing isolation between containers. 
Our Docker Image (based on official Dockerhub CentOS 7.3 and gcc 4.8) was created only once, 
and it included the {\tt ctools} suite with its proper dependencies 
(Gammalib, Python and C$++$ libraries), and the  calibration IRFs files 
\citep[see details on requirements in][]{Gammalib_ctools_2016}. 

\subsection{AWS Elastic Cloud Computing (EC2) } \label{methods:ec2}

Amazon Elastic Cloud Computing (EC2)\footnote{\href{https://aws.amazon.com/ec2/}{https://aws.amazon.com/ec2/}. }  
is a cloud service that offers computational power through 
objects called ``instances''. 
We can consider an instance as a computer node connected to the Internet,  
with its own CPU, RAM, and disk space  (called instance storage), as shown in Fig.~\ref{methods:fig:awsec2}. 
These resources are fully maintained in the AWS Availability Zones, that are large server farms distributed across the world.
One of the main advantages, in addition to the large available computational power, is that instances are fully maintained
by AWS and the final user does not need to care about hardware configuration, maintenance, and general housekeeping. 
Moreover, the underlying hardware is regularly upgraded without any disruption of the service or user required actions. 

Instances are acquired and fired up at need, depending on the number of realizations to be performed.  
When their work is completed instances are switched off, and all their content, including the instance storage, deleted. 
The AWS platform offers many different types of instances with various configurations in terms of 
CPU, RAM, and network performance. 
For our purpose, we chose spot instances (which are basically an unsold computation capacity 
supplied at very low prices) of type ``m4x16large'' that 
offer 64 Virtual Central Processing Units (vCPUs) and 256 GB of RAM, thus sufficient to run at least 62 independent realizations 
part of a {\tt ctools} simulation (considering 2 vCPU for the underlying operating system and Docker hypervisor).
We decided to opt for ``m4x16large' EC2 instance type since it combines a reasonable ratio of vCPU and RAM while offering a best compromise between availability, when required as Spot, and price. Our EC2 cluster  configuration here explained reaches a tradeoff between number of running instances, cost per hour and number of Elastic Block Storage volumes (hard disk equivalent) deployed in the region.
We generally fired up a cluster of 30 of such instances so that 
we could run up to 1860 simultaneous realizations.

\subsection{Amazon Simple Storage Service (S3) } \label{methods:s3} 

AWS Simple Storage Service (S3)\footnote{\href{https://aws.amazon.com/S3/}{https://aws.amazon.com/S3/}. }  
is a service used to save large amounts of data with high 
consistency and durability, while affording a very fast access time.  
It resembles the commonly used file storage services but it  also offers 
advanced functionalities that allowed us  to coordinate other cloud-based facilities 
(like Lambda function, see Sect.~\ref{methods:lambda}). 
We used S3 to store data (input XML files and scripts required to run realizations,
output FITS event files, and generally all results from our realizations) 
produced by the cluster of instances before they were switched off
(see Sect.~\ref{methods:ec2}). 
We also used this service as a long term storage 
by moving old data to a particular storage class called 
{\tt Glacier}\footnote{\href{https://aws.amazon.com/glacier/}{https://aws.amazon.com/glacier/}. }  
that has lower maintenance costs and increased long-term durability of data. 
For this project we needed less than 1\,GB  to store all the final results.

\subsection{Amazon Lambda   } \label{methods:lambda}
{\tt Lambda}\footnote{\href{https://aws.amazon.com/lambda/}{https://aws.amazon.com/lambda/}. }   is a computer service that runs functions in response to some events triggered 
in the cloud (for example, a file upload to S3). 
The resources required to run functions and the triggers are automatically managed. 
We used AWS {\tt Lambda} to coordinate the step between upload of simulations inputs to S3 
and the consequent EC2 cluster fire-up.

\begin{figure}[t] 
\hspace{-1.8truecm}
\includegraphics[scale=0.40,angle=-90]{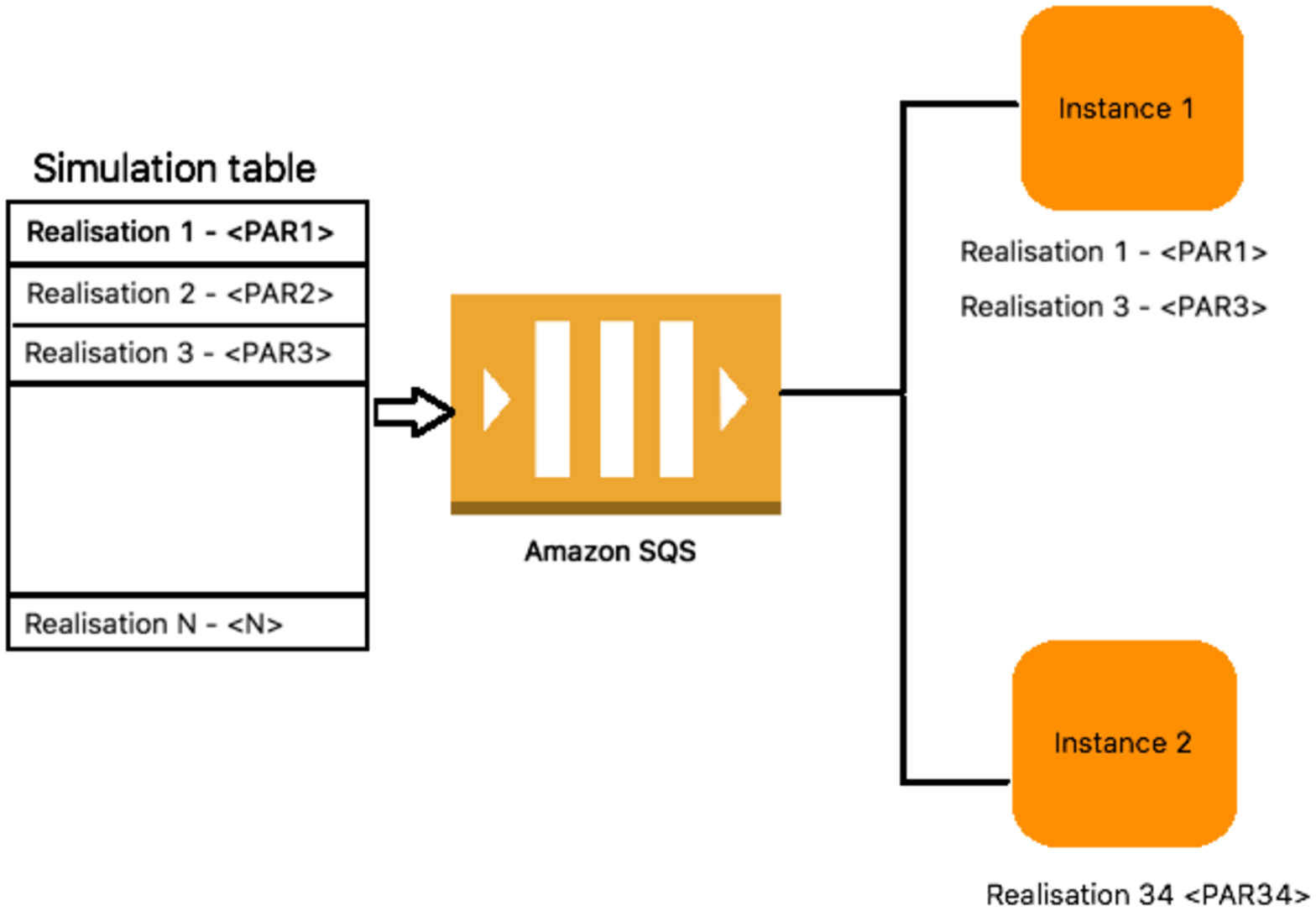}
\vspace{-0.80truecm}
\caption{AWS SQS and AWS EC2 interfaces for the distribution of the workload. 
At the beginning, the table is pushed to SQS and, runtime, 
each  instance receives some rows of realization that are performed using the {\tt ctools}  
suite containerized through {\tt Docker}.
} 
\label{methods:fig:sqs} 
\end{figure} 

\begin{figure*}[t] 
\vspace{-0.5truecm}
\hspace{-0.5truecm}
\includegraphics[scale=.60,angle=90]{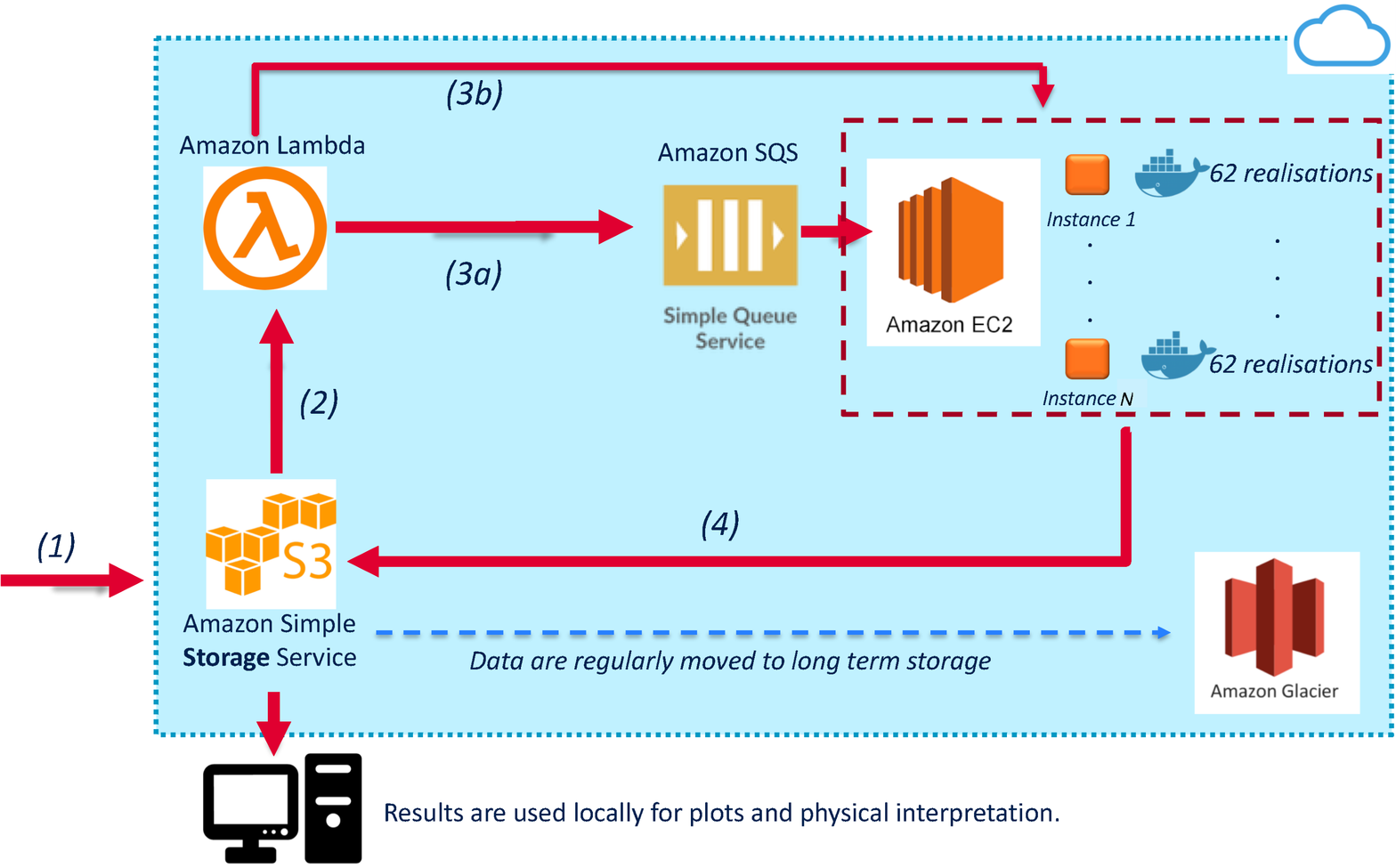}
\caption{
The cloud-based AWS architecture for parallel and distributed CTA simulations. 
This solution foresees the implementation of Amazon EC2 Service as computational power and 
Amazon S3 and Glacier as baseline for the required storage. The Amazon Lambda and Amazon 
SQS services orchestrate the logical workflow of the system 
(see Sect.~\ref{methods:architecture} for details). 
} 
\label{fig:archaws} 
\end{figure*} 

%
%
 \begin{table*} 	
 \tabcolsep 3pt  
\begin{center} 	
 \caption{Mean values and 1-$\sigma$ uncertainties based on $N=1000$ realizations. 
Input values for the Crab spectrum are $k_0=5.7\times10^{-16}$\,ph\,cm$^{-2}$\,s$^{-1}$\,MeV$^{-1}$, 
$\gamma=-2.48$; the expected 0.1--100\,TeV band flux is $5.248\times10^{-10}$\,erg\,cm$^{-2}$\,s$^{-1}$. 	
For the other sources a scaling factor of 0.1, 0.01, and 0.001 to $k_0$ is applied. }
 \label{methods:tab:results} 
\begin{tabular}{lrcccccc}	
 \hline 
 \hline 
 \noalign{\smallskip} 
  Model        &Expo. & $k_0$	                                 & $\gamma$	  	   & TS		 	            &  Events		     &   Flux &  $\Delta$F/F\\
                   & (h)     & (ph\,cm$^{-2}$\,s$^{-1}$\,MeV$^{-1}$)	&                 &           		            & 	                     &   (erg\,cm$^{-2}$\,s$^{-1}$)  &   \\ %
 \noalign{\smallskip} 
 \hline 
 \noalign{\smallskip} 
  Crab	                            &5  & $(5.70\pm 0.03)\times10^{-16}$ & $2.480\pm0.005$  & $227\,288\pm1\,661$  &$36\,833\pm201$ & $(5.248\pm 0.036)\times10^{-10} $ & 0.0069 \\ 
0.1Crab	                            &5   & $(5.70\pm0.11)\times10^{-17}$  & $2.479\pm0.016$  & $10\,213\pm291$         & $3\,684\pm60$    & $(5.26\pm 0.12) \times10^{-11}$ & 0.022 \\ 
0.01Crab	                            &5   & $(5.74\pm0.60)\times10^{-18}$  & $2.483\pm0.072$  & $301\pm42$                  & $368\pm20$        & $(5.26\pm 0.44) \times10^{-12}$ & 0.084 \\  
mCrab\tablenotemark{a}      &5   & $(6.8\pm4.6)\times10^{-19}$  & $2.6\pm1.2$  & $9\pm7$                        &  $37\pm6$           & $(5.29\pm 3.5)\times10^{-13}$      & 0.67\\  
mCrab\tablenotemark{a}      &10 & $(6.3\pm3.3)\times10^{-19}$ & $2.50\pm0.65$ & $14\pm8$                      &  $73\pm9$           & $ (5.34\pm2.2)\times10^{-13}$      &  0.40 \\   
mCrab                                  &50 & $(5.9\pm1.4)\times10^{-19}$  & $2.49\pm0.13$  & $68\pm18$                    & $328\pm18$        & $(5.30\pm 0.83) \times10^{-13}$ & 0.16 \\  
 \noalign{\smallskip}
 \noalign{\smallskip}
  \hline
  \end{tabular}
\tablenotetext{a}{The source is not detected in most of the realizations, see Fig.~\ref{methods:fig:mCrab_over_TS}. }
  \end{center}
  \end{table*} 

\subsection{Amazon Simple Queue Service (SQS)  }  \label{methods:sqs}
As we have seen, a simulation is made of a thousand independent realizations, 
each with its own input parameters, so that a simulation could be thought as a 
table where each row contains the information to run a single realization 
(see Fig.~\ref{methods:fig:sqs}). 
In this scenario, Amazon Simple Queue Service (SQS)\footnote{\href{https://aws.amazon.com/sqs/}{https://aws.amazon.com/sqs/}. } 
 allows us to store this table and distribute rows 
(technically called ``messages'') across the cluster in the common producer-consumer paradigm. 
In our approach, the producer is the simulation table (stored only at the beginning of the whole simulation) 
with its large list of rows, while consumers are CPUs distributed across the EC2 cluster 
(see Fig.~\ref{methods:fig:sqs}). Each of them is in charge of completing one full realization. 
The possibility to exploit SQS provides a method to host 
queued messages and reduces connectivity problem between consumers.

\subsection{The AWS-based cloud architecture } \label{methods:wholepicture}
As we described the concepts and services required to run a full simulation 
we are now able to detail the flow of data on the AWS architecture implemented 
for CTA (see Fig.~\ref{fig:archaws}).

The distributed computation starts by uploading in S3 (step 1) a tarball file 
containing the XML file description of the source to simulate 
and the simulation table (as plain text file). 
As soon as the upload is completed a trigger is raised (step 2) and an Amazon Lambda function 
pushes the simulation table 
into the distributed and highly available first-in first-out queue SQS (step 3a). 
Then, it
creates a homogenous set of EC2 instances and fires them up, 
with the number of instances varying accordingly to the overall size of the simulation (step 3b). 
This allows the distribution of realizations among many computers. 
When each EC2 instance is online, it automatically pulls up to 62 messages from the SQS queue. 
Each one of these messages contains the information necessary to run the realization 
on the node through execution of the Docker containers.  
Then, the EC2 instance waits for all containers to finish their computation 
and the outputs, temporarily saved on the local instance storage,  
are saved onto S3 (step 4).  
Processed data can then be recovered from S3 for subsequent analysis and physical 
 interpretation of the results.

          \section{Results } \label{methods:results}

%
%
%
\begin{figure}[t] 
\includegraphics[scale=.50]{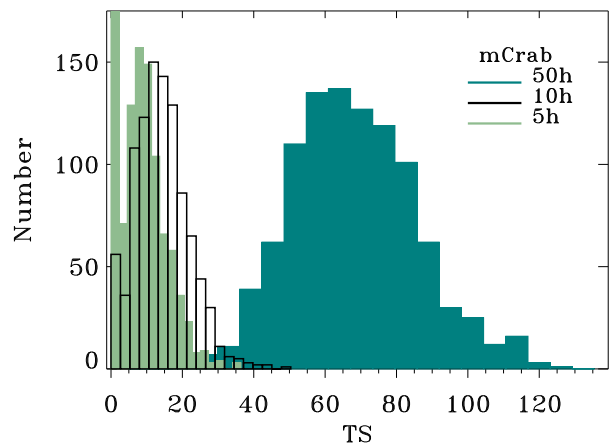}
\vspace{0.3truecm}
\caption{Comparison of the distributions of the test statistic TS values 
for the mCrab source as a function of exposure time (5\,h, 10\,h, and 50\,h). 
The source is not detected in most realizations for 5\,h and 10\,h and the 
TS distribution cannot be modelled by a Gaussian. 
} 
\label{methods:fig:mCrab_over_TS} 
\end{figure} 

\begin{figure} 
\includegraphics[scale=0.50]{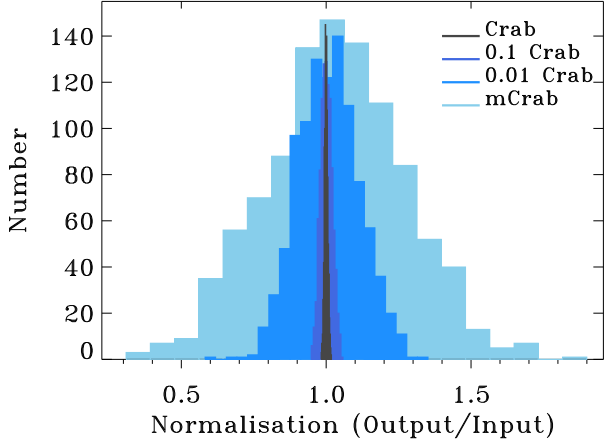}
\vspace{0.3truecm}
\caption{Distributions of the power-law fit normalisation (Col.~3 of Table~\ref{methods:tab:results}), 
as a function of input flux. All distributions have been normalised to the input flux (see Sect.~\ref{methods:sims}): 
1\,Crab (charcoal, in units of $5.7\times10^{-16}$\,ph\,cm$^{-2}$\,s$^{-1}$\,MeV$^{-1}$);  
0.1\,Crab (dodger blue, in units of $5.7\times10^{-17}$\,ph\,cm$^{-2}$\,s$^{-1}$\,MeV$^{-1}$); 
0.01\,Crab (navy blue, in units of $5.7\times10^{-18}$)\,ph\,cm$^{-2}$\,s$^{-1}$\,MeV$^{-1}$); 
1\,mCrab (sky blue, in units of $5.7\times10^{-19}$\,ph\,cm$^{-2}$\,s$^{-1}$\,MeV$^{-1}$). 
Exposure times are 5\,h for all but mCrab case, where 50\,h were used. 
} 
\label{methods:fig:distro_kappa4} 
\end{figure} 
\begin{figure} 

\includegraphics[scale=.50]{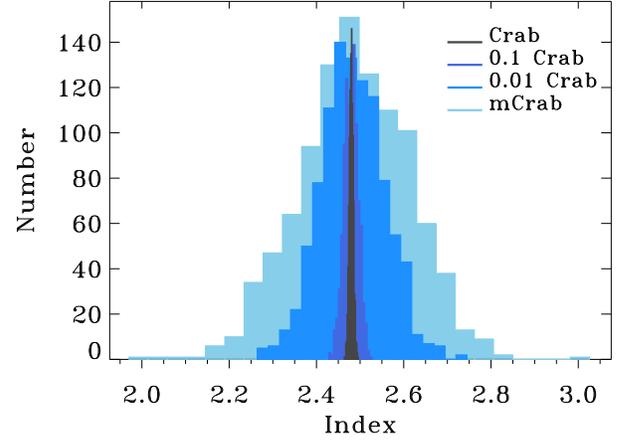}
\vspace{0.3truecm}
\caption{Distributions of the index in the power-law fit, as a function of input flux 
(Col.~4 of Table~\ref{methods:tab:results}).  
} 
\label{methods:fig:distro_gamma4} 
\end{figure} 
\begin{figure}[t] 
\hspace{-0.3truecm}
\includegraphics[scale=0.45]{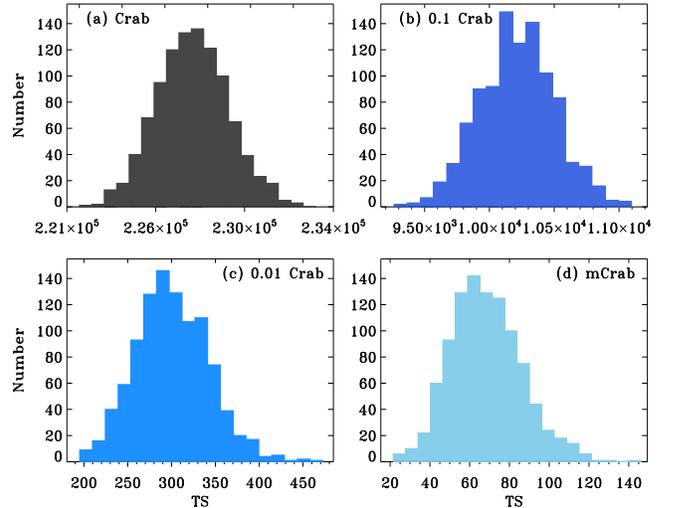}
\caption{Distributions of the TS, as a function of input flux 
(Col.~5 of Table~\ref{methods:tab:results}), 
(a): 1\,Crab; (b):  0.1\,Crab; (c):   0.01\,Crab; (d): 1\,mCrab. 
} 
\label{methods:fig:distro_ts4} 
\end{figure} 
\begin{figure}[t] 
\includegraphics[scale=0.50]{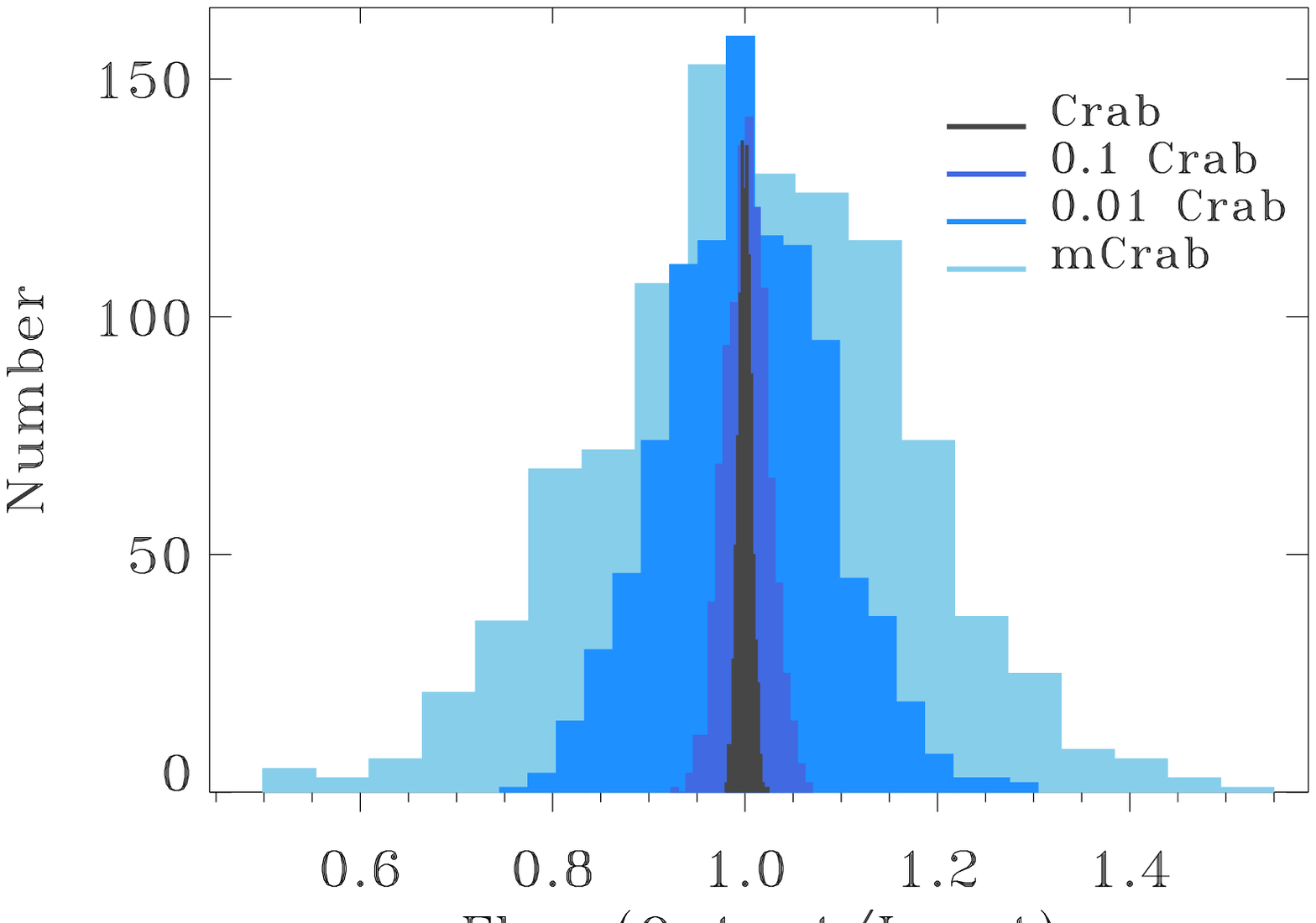}
\vspace{0.3truecm}
\caption{Distributions of the fluxes (Col.~7 of Table~\ref{methods:tab:results}), 
 normalised to the input flux: 
1\,Crab (in units of $5.248\times10^{-10}$\,erg\,cm$^{-2}$\,s$^{-1}$);  
0.1\,Crab (in units of  $5.248\times10^{-11}$\,erg\,cm$^{-2}$\,s$^{-1}$); 
0.01\,Crab (in units of  $5.248\times10^{-12}$\,erg\,cm$^{-2}$\,s$^{-1}$); 
1\,mCrab (in units of  $5.248\times10^{-13}$\,erg\,cm$^{-2}$\,s$^{-1}$). 
} 
\label{methods:fig:distro_flux4} 
\end{figure} 

In Table~\ref{methods:tab:results} we report the mean values and 1-$\sigma$ uncertainties 
of the simulation parameters obtained with $N=1000$ independent realizations 
for the 4 input model fluxes. We consider a source detected with 
a high significance when $TS \geq 25$  \citep[][]{Mattox1996:Egret} 
and a low significance when $10 \leq TS < 25$. 
The source will not be considered detected for $TS <10$.
We note that the mCrab source is not detected in all realizations
for a 5\,h and 10\,h exposure times. 
This is graphically shown in the distributions of TS values  
in Fig.~\ref{methods:fig:mCrab_over_TS}. 
Therefore, we shall only consider the case of 50\,h 
for the mCrab source and the 
TS distribution cannot be modelled by a Gaussian. 
For each of the 4 input model fluxes, we show the distributions of the 
normalisations (Fig.~\ref{methods:fig:distro_kappa4}), 
spectral indices (Fig.~\ref{methods:fig:distro_gamma4}), 
TS values (Fig.~\ref{methods:fig:distro_ts4}), and 
derived fluxes in the 0.1--100\,TeV energy band (Fig.~\ref{methods:fig:distro_flux4}).

As expected, the spread in the parameter values depends strongly on the 
input flux. Indeed, we can clearly see that the parameters are progressively 
more constrained as the input flux increases, 
so that the relative uncertainty on the output flux decreases down to $\sim16$\% for a mCrab and to within $\sim0.7$\% for a Crab
(see Table~\ref{methods:tab:results}). 

In other terms, Figs.~\ref{methods:fig:distro_kappa4}--\ref{methods:fig:distro_flux4} 
show how, due to the fact that individual realizations vary considerably 
in terms of derived parameter values and their statistical uncertainties 
(see in particular for the mCrab case, where the effect is more evident), 
a large number of realizations is always required in order to obtain 
average output parameters that are truly representative of the input spectrum. 
This is particularly critical for faint sources, since the background 
can become dominant. 
This in turn implies longer and longer run times for simulations of progressively
fainter sources.

          \section{Budget discussion } \label{methods:budget}

As reported in Table~\ref{methods:tab:crab}, the typical run time (single core) 
for a total of  $N=1000$ independent realizations 
varied from 8\,hr (a Crab for 5\,h exposure time) to over 3\,days 
(a mCrab, 50\,h exposure time).  
We further note that our test sources required a  considerably simpler treatment 
than simulations of realistic astrophysical sources and lines of investigation, 
for which several more factors need to be taken into account. 

The {\it Galactic diffuse background} is dominant for all sources close to the
        Galactic plane (e.g.\ ($| b|<10^{\circ}$).  
        Based on our simulations performed on HESS~J0632$+$057 (M.\ Chernyakova et al., in prep),
        the run time can increase up to a factor of 5--10. 
        Detailed simulations, including both event generation and likelihood analysis,   
        of a typical astrophysical Galactic source can then take months of CPU time. 

The {\it application of the energy dispersion matrix} for sources requiring a detailed 
                         spectral treatment in the softer energies (e.g.\ E$\la 500$\,GeV) 
                         and will be shown 
                         (\citealt{Romano2018}; 
                         \citealt{Lamastra2019})
                         to increase the run time by a factor of 5--10.

VHE astrophysicists are now also tackling the task of simulating a {\it large populations of sources}, 
       such as e.g.\ active galactic nuclei, gamma-ray bursts \citep[][]{Ghirlanda2015:popGRB}, binary sources. 
       This task can quickly become prohibitive on less than a cluster of high-performance machines. 
Moreover, in all cases the presence of contaminating sources in the 5\,deg FOV considered 
       for event extraction, a likely case in crowded regions, 
       will scale the run time with the number of contaminating sources.

A further complicating issue is the storage of intermediate data and final data. 
For our four test sources, the typical event file (the largest intermediate product of our 
pipeline) size was 22\,MB for 5\,hr exposure, with the mCrab case reaching $\sim160$\,MB.
The test case as a whole required a total storage space of 0.5\,TB on a fast access disk. 
Realistic projects our group is currently pursuing are producing $\sim300$\,MB event files, 
and dozens of TB of intermediate products. For the average standalone machine this implies 
regular backing-up and storage of final products and deletion of unused intermediate products.

As a conclusion, we report a comparison of the costs required for running our test case 
on a local machine and on AWS.  
Our local machine (Intel\textsuperscript{\textregistered}  Xeon\textsuperscript{\textregistered}  CPU E5-2620)
cost about 7000\,USD, considering both HD storage and uninterruptible power supply (UPS),
which, assuming a mean time between failures (MTBF) of three years, 
implies an hourly cost of 0.022 USD\,hr$^{-1}$\,core$^{-1}$. 
The test case therefore cost about 4\,USD, in terms of CPU, and took about 172 core hours to run.
Table~\ref{methods:tab:costs} shows (Col.\ 2, Rows.\ 1--4) the 
costs for running the simulations locally, 
storage, and maintenance, respectively. 
These costs do not include the costs that the project needs to sustain for power and air conditioning, or 
data management (creating queues for script running, cleaning-up storage disks, and general 
housekeeping) during the simulations, generally in terms of human effort and permanent storage.

In comparison, on average AWS spot instances in the cheapest availability zones, 
cost about about 0.0078\,USD\,hr$^{-1}$ core$^{-1}$, 
so that the test case cost about 1.3\,USD for the run and 5\,USD for storage 
(see Table~\ref{methods:tab:costs}, Col.\ 3, Rows.\ 1--4). 
This implies that use of AWS reduced the hourly cost per vCPU core by a factor of $\sim$ 3. 
The test case run was also performed successfully on AWS using our architecture 
based on a cluster of 60 m4.16xlarge spot instances 
in only $\sim$ 0.5\,hrs, which is a factor of 350 faster than locally.  
The combination of burst simulations (long but infrequent) and the possibility to 
exploit the unsold capacity on AWS cloud platform allows us to both reduce the required 
amount of time to run a full simulation while guaranteeing low cost being billed for just the used CPU time. In this way, it is not necessary to pay in advance for a sitting idle local system.
 We note that {\tt ctools} suite could be run 
using the in-memory pipeline to avoid I/O  for temporary disk data storage of event files, further decreasing the overall cost. 
Clearly, for a small project such as the test case we presented  
a factor of 3 reduction in CPU cost may not go a long way to justify the 
time spent to make the codes parallel and distributable and to learn to use 
the cloud services. 
Nor a factor of a few hundred in run time, if time is not of the essence. 
However, as Table~\ref{methods:tab:costs} shows, other factors need to be 
taken into account when dealing with simulations of real astrophysical sources. 
While on AWS all intermediate processes  
are taken care by the Lambda functions, 
when running simulations locally, the human effort required
increases with the number of realizations being run. 
While it is indeed true that it is feasible to do many things with ctools without having a computer farm, the determination of the TS distribution to assess source detectability (for which a large number of realization is mandatory) require a considerable amount of computational power.  This applies also to the simulation of large volumes of data or a complex analysis pipeline, such as needed for building the GPS catalogue. In these end-to-end full simulations months of CPU time could be required, achievable with our proposed concept for a small amount of money and 
fully scalable in relation with the complexity of projects.

  \begin{table} 	
 \tabcolsep 8pt  
\begin{center} 	
 \caption{Budget of {\tt ctools} simulations for the test case (Sect.~\ref{methods:sims}). 	
 \label{methods:tab:costs} }	
 \begin{tabular}{lll}	
 \hline 
 \hline 
 \noalign{\smallskip} 
Costs  (USD)                & Local	          & AWS    \\
  \noalign{\smallskip} 
 \hline 
 \noalign{\smallskip} 
Simulations         & 4                &   1.3 \\ 
Storage                &  2.0\tablenotemark{a}   &   5.0\tablenotemark{b}   \\ 
Maintanence         &  2.5                 &   --   \\ 
 \noalign{\smallskip} 
Total                   &  8.5                &   6.3   \\ 
\noalign{\smallskip} 
 
\hline 
 \noalign{\smallskip} 
     Run Times (hr)  & 172       &  0.5 \\ 
 \noalign{\smallskip} 
 \noalign{\smallskip} 
\hline 
\hline 
\noalign{\smallskip} 
      Scaling Factors                              & Local	 & AWS    \\
 \noalign{\smallskip} 
 \hline 
 \noalign{\smallskip} 
Galactic Background &     $\times$5--10     & $\times$5--10          \\
Energy dispersion     &     $\times$5--10     & $\times$5--10          \\
   \noalign{\smallskip}
  \hline
  \end{tabular}
\tablenotetext{a}{Only considering temporary storage on HDs, that need to be backed-up for the  
storage of final products, and cleaned periodically from unused intermediate products.}
\tablenotetext{b}{Including long term storage and billing costs.} 
  \end{center}
  \end{table} 

\vskip 0.5in
\section*{acknowledgements}

We are grateful to the referee for his/hew review that improved the quality of our manuscript. We thank L.\ Foschini, J. Bregeon, G Maier and K. Kosack for helpful discussions. \\ 

%
The authors acknowledge contribution from the grant INAF CTA--SKA, 
``Probing particle acceleration and $\gamma$-ray propagation with CTA and its precursors'' (PI F.\ Tavecchio). 

%
This research has made use of the CTA instrument response functions provided by the CTA Consortium and Observatory, 
see \href{https://www.cta-observatory.org/science/cta-performance/}{https://www.cta-observatory.org/science/cta-performance/} 
(version prod3b-v1) for more details. \\ 

%
This research made use of ctools, a community-developed analysis package for Imaging Air Cherenkov Telescope data. 
ctools is based on GammaLib, a community-developed toolbox for the high-level analysis of astronomical gamma-ray data. \\

%

%
         We gratefully acknowledge financial support from the agencies and organizations 
          listed here: 
\href{http://www.cta-observatory.org/consortium_acknowledgments}{http://www.cta-observatory.org/consortium\_acknowledgments}
%
This paper has gone through internal review by the CTA Consortium.  \\ 

\facilities{Cherenkov Telescope Array}
\facilities{CTA}
\software{ctools \citep{Gammalib_ctools_2016}} 

\bibliographystyle{aasjournal} 

\end{document}